\documentclass{iaus}
\usepackage{graphicx}

\title[Searching for Wolf-Rayet Stars in M101] %% give here short title %%
{Searching for Wolf-Rayet Stars in M101}

\author[J. L. Bibby et al.]   %% give here short author list %%
{J. L. Bibby$^1$, P.A. Crowther$^2$, 
  A.F.J. Moffat$^{3}$, M.M. Shara$^{1}$,\\  D. Zurek$^{1}$ \and L.Drissen$^{3}$}
\affiliation{$^1$Dept. of
  Astrophysics, American Museum of Natural History \\ Central Park
  West @ 79th St, New York, NY 10024, USA \\ email: {\tt jbibby@amnh.org} \\[\affilskip]
  $^2$Dept. of Physics \& Astronomy, University of Sheffield,  \\
  Hounsfield Rd, Sheffield, S3 7RH, UK  \\ $^{3}$ D\'{e}pt. de physique,
  Universit\'{e} de Montr\'{e}al, C.P. 6128, \\Succursale Centre-Ville,
  Montr\'{e}al, QC, H3C 3J7, Canada }

\pubyear{2012}
\volume{279}  %% insert here IAU Symposium No.
\pagerange{xxx--yyy}
\jname{Death of Massive Stars: Supernovae and Gamma-Ray Bursts}
\editors{P. Roming, N. Kawai \& E. Pian, eds.}
\begin{document}

\maketitle

\begin{abstract}
  Wolf-Rayet (WR) stars are the evolved descendants of massive O-type
  stars and are considered to be progenitor candidates for Type Ib/c
  core-collapse supernovae (SNe).  Recent results of our HST/WFC3 survey of
  Wolf-Rayet stars in M101 are summarised based on the detection
  efficiency of narrow-band optical imaging compared to broad-band methods. We
  show that on average of 42\% WR stars, increasing to $\sim$85\%
  in central regions, are \textit{only} detected in the narrow-band
  imaging. Hence, the non-detection of a WR star at the location of $\sim$10
  Type Ib/c SNe in broad-band imaging is no longer strong evidence for a
  non-WR progenitor channel.

\keywords{stars: Wolf-Rayet, galaxies: M101, Survey.}
\end{abstract}

\firstsection % if your document starts with a section,
\section{Introduction}
Wolf-Rayet (WR) stars are evolved massive O--type stars which are
predicted to be the progenitors of Type Ibc core-collapse supernovae
(ccSNe). However, to date there has been no direct confirmation of the
WR-SNe connection. Pre-SNe broad-band images have failed to reveal
the progenitor of $\sim$10 Type Ib/c SNe \cite[(Smartt
2009)]{Smartt2009}.

WR stars can be classified into two main subtypes, nitrogen--rich (WN)
and carbon--rich (WC) stars, which reveal the products of CNO burning
and triple alpha reactions, respectively. WR stars exhibit a unique
emission line spectrum which is dominated by He\,{\sc ii}$\lambda$4686
emission lines for WN stars while WC spectra, which also show weaker
He\,{\sc ii}$\lambda$4686, are dominated by C\,{\sc iii}$\lambda$4650
and C\,{\sc iv}$\lambda$5808 \cite[(Crowther
2007)]{Crowther2007}. These strong emission lines can be
0.2-2.5\,magnitudes brighter than the adjacent continuum, making WR
stars easy to detect using narrow-band imaging techniques.

Type Ib SNe are hydrogen--poor, while Type Ic SNe are both hydrogen--
and helium--poor. The similarities between the observed Type Ib and Ic
SN spectra and the chemical composition of the WN and WC stars makes
them strong progenitor candidates, respectively.

\section{M101}
M101 is a grand-design spiral galaxy which lies face-on at a distance
of 6.2\,Mpc \cite[(Shappee \& Stanek, 2011)]{Shappee2011}. It has a
high star-formation rate of at least $\sim$4.5\,M$_{\odot}$yr$^{-1}$
based on the H$\alpha$ flux or $\sim$6.8\,M$_{\odot}$yr$^{-1}$ from
far UV imaging \cite[(Lee et al. 2009)]{Lee2009}. 

% M101 recently hosted the Type Ia SN 2011fe, for which the progenitor
% was constrained by \cite[Li et al. (2011)]{Li2011} using archival
% HST images.

There is a wealth of HST/ACS archival data for M101 which was obtained
under the legacy program, however the most effective way to identify
WR stars is via narrow-band imaging techniques centered on the WR
emission lines. We obtained 18 pointings of M101 using WFC3/F469N over
36 orbits in cycle 17 (PI: Shara).

Identifying WR stars beyond the Local Group with ground-based imaging
is challenging since we can only resolve them on scales of $\sim$25pc. The
emission lines of WR stars, although strong, can easily be diluted
by strong continuum from other OB stars in the unresolved region
\cite[(Bibby \& Crowther 2010)]{Bibby2010}. The high spatial
resolution images allow us to resolve sources down to $\sim$3\,pc,
decreasing the contamination of the WR emission by continuum sources.

Typically WR stars span an absolute magnitude range of M$_{V}$\,=\,--4
to --8\,mag; however, our ground-based studies do not extend fainter
than M$_{V}$\,=\,--5\,mag. The improved sensitivity of HST allows us
to identify the faintest and least massive WR stars, detecting stars
to M$_{F469N}$\,=\,--3.5\,mag and M$_{F435W}$\,=\,--4\,mag.

\section{Identifying candidates}
The ACS and WFC3 data were re-drizzled onto the same scale of 0.05
arcsec pix$^{-1}$ using \textsc{multidrizzle}. Photometry was
performed with the stand-alone \textsc{daophot} package.  Stars that
had at least a 3 sigma excess in the F469N filter compared to the
F435W and F555W filters were identified as WR candidates. These
candidates were then visually inspected using the ``blinking'' method
\cite[(Moffat \& Shara, 1983)]{Moffat1983}.

%The F435W image was scaled, based on bandwidth and
%transmission efficiency, to act as a pseudo narrow-band continuum and
%was then subtracted from the F469N image, producing a``net'' image
%which identified regions of He\,{\sc ii} excess.

\begin{figure}
\centering
\includegraphics[width=0.65\columnwidth, angle=-90]{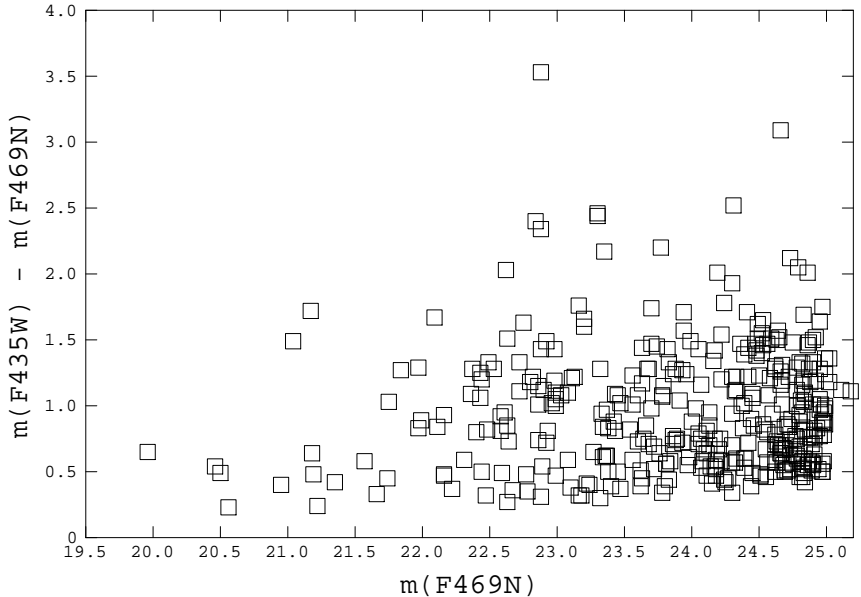}
\caption{F469N magnitude versus excess magnitude relative to the F435W
  image for the 372 WR candidates identified in four of the HST/WFC3
  pointings.}
\label{magnitudes}
\end{figure}

Four of the WFC3/F469N pointings have currently undergone analysis and
have revealed 372 WR candidates with a He\,{\sc ii} $\lambda$4686
excess of at least 3 sigma (Fig. \ref{magnitudes}). The brightest
candidates with m(F435W)-m(F469N) excess of $\leq$2.5\,mag are likely
to be single WR stars, whereas the fainter candidates at
m(F469N)$\sim$25\,mag with small m(F435W)-m(F469N) excesses
$\leq$0.6\,mag are more likely to host multiple WR stars in unresolved
clusters.

\begin{figure}
\centering
\includegraphics[width=0.32\columnwidth, angle=-90]{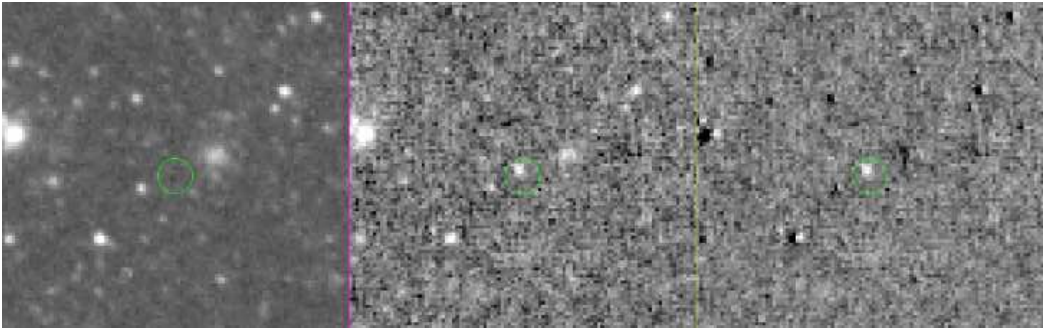}
\caption{Postage stamp image ( $\sim$30$\times$30 arcsec) of a WR
  candidate in M101. The F435W broad-band continuum image (left) and
  narrow-band F469N image (centre) are subtracted to produce the
  ``net'' image (right), revealing He\,{\sc ii}$\lambda$4686 emission
  indicating the presence of a WR star.}
\label{blink}
\end{figure}

From the photometric analysis we were also able to identify sources in
the F469N image that were not identified in either the F435W or F555W
images. This is indicative of a faint WR star with a continuum which
lies below the detection threshold of the broad-band images. We
identified an additional 269 WR candidates that were only detected in
the narrow-band images, which were again checked using the blinking
method (Fig.\ref{blink}).

We are currently awaiting execution of follow-up multi-object
spectroscopy with Gemini-North/GMOS for a sample of these candidates
($\sim$35\%) which will allow us to (i) determine the multiplicity of
the sources, (ii) assign a spectral classification of WN or WC to
the candidate and (iii) infer the subtype of the remaining candidates
for which we did not obtain spectroscopic confirmation.

%\section{WR stars in broad-band images}

\section{Conclusions \& Future Work}
We identified 641 WR candidates within four HST/WFC3 pointing of M101,
42\% of which are \textit{only} detected in the narrow-band
image. This is much higher than the 25\% found from ground-based
studies of other nearby spiral galaxies, e.g. \cite[Bibby \& Crowther
(2012)]{Bibby2012}, which is most likely due to the improved
sensitivity and spatial resolution of HST which allows us to detect
fainter WR stars. This work highlights the effectiveness of high,
spatial resolution narrow-band observations in detecting WR stars. If
we hope to confirm or, equally important, rule out WR stars as the
progenitors of Type Ib/c SNe, then we require a \textit{complete}
sample of WR stars in several nearby galaxies.

% If we consider the four pointings individually the fraction of WR
% stars not detected in the broad-band images ranges from 25\% in the
% outer regions increasing to 85\% towards the centre.
\vspace{0.2cm}
\noindent Once analysis of M101 is complete we will be able to;

(i) Investigate how the number of WR stars varies with the number of O
stars and Red Supergiants (RSG) across the galaxy as a function of
metallicity and compare the results to predictions from theoretical
evolutionary models.

(ii) Degrade our HST imaging and re-run the analysis to determine how
many WR stars are not detected, and hence estimate the true
completeness of our ground--based surveys.

(iii) Analyse the properties of the regions associated with WR stars
to assess whether they are consistent with Type Ib/c SNe.

(iv) Expand our existing catalogue of WR stars which can then be
referred to in the event that a Type Ib/c SNe occurs in one of these
galaxies. This will allow strong, direct observational evidence for,
or against, the WR-SNe connection.

\begin{discussion}

\discuss{J. Shiode}{What contaminating sources might masquerade as WR
  stars? }

\discuss{J. Bibby}{We worried about contamination from planetary
  nebulae, however our detection limits are not deep enough to detect
  PNe.}

\discuss{J. Shiode}{Can evolutionary calculations really be considered
predictions for N(WR)/N(O) given the uncertainty in very massive star
mass-loss histories?}

\discuss{J. Bibby}{N(WR)/N(O) ratios have been relatively consistent
  with observations where we have a complete WR sample.  We should not
  blame the models until we are sure we are detecting all of the WR
  population. Our M101 survey will detect the faintest WR stars hence
  our ratios should be reliable and so if they don't match predictions
  it gives theoreticians something to work with.}

\discuss{S. Chakraborti}{What if some of the stars are just variables?}

\discuss{J.Bibby}{We have not encountered such a problem in our
  ground-based surveys but if O stars can vary by up to a magnitude on
  timescales of a few years then this is something we will need to
  look into. If our follow-up spectroscopy is contaminated by such
  stars then clearly it is a problem and we need to look at the
  properties of these stars to see if we can distinguish them
  from our WR candidates.}

\discuss{T. L. Astraatmadja}{Why not look for the WR stars in our
  Galaxy?}

\discuss{J. Bibby}{There are indeed ongoing WR surveys of the Milky
  Way. However due to the dust extinction in our galaxy IR rather than
  optical techniques are used. The overall aim of our work is to identify
  a future SNe progenitor, so surveying many nearby galaxies increases
  our chances. We select our galaxies based on several criteria, one
  of which is a face-on orientation to avoid the problem of high
  extinction.}

\discuss{T. L. Astraatmadja}{Do WR stars in other galaxies have the
  same characteristics?}

\discuss{J. Bibby}{Yes, they still have strong emission lines, however the
  strength of the lines vary with the metallicity of the environment,
  i.e. in a low metallicity environment the line flux will be reduced
  as was demonstrated by \cite[Crowther \& Hadfield
  (2006)]{CrowtherHadfield2006}.}

\discuss{P. Mazzali}{Can you distinguish from binary WR stars in your
  survey?}

\discuss{J. Bibby}{Unfortunately we only obtain one epoch of spectra
  for a sample of WR stars so we cannot identify which stars are in a
  binary system from radial velocity motion.}

\end{discussion}

\end{document}